\documentclass[review]{elsarticle}
\usepackage{lineno,hyperref,multicol}
\usepackage[top=35truemm,bottom=30truemm,left=25truemm,right=25truemm]{geometry}
\modulolinenumbers[5]
 
\usepackage{amsmath}
\usepackage{subfigure}

\begin{document}
\normalsize
\begin{frontmatter}

\title{\LARGE\bf Measurement of dE/dx resolution of TPC prototype \\ with gating GEM exposed to an electron beam}

\author{\rm \large Aiko Shoji${}^\text{*}$\footnote{Talk presented at the International Workshop on Future Linear Colliders (LCWS2017), Strasbourg, France, 23-27 October 2017. C17-10-23.2.\\E-mail address: t5617002@iwate-u.ac.jp}\\ on behalf of the LCTPC group}
\address
{ \normalsize ${}^\text{*}$Iwate University, Department of Electrical and Electronic Engineering, \\4-3-5 Ueda, Morioka, Iwate 020-8551, Japan}

\begin{abstract}
The Time Projection Chamber (TPC) can identify particle species using the energy loss (dE/dx) measured by the readout pad rows. The dE/dx resolution for the ILD (International Large Detector) -TPC for the International Linear Collider experiment is expected to have high resolution for clear identification. The dE/dx resolution of our prototype TPC with gating GEM to suppress ions feedback was measured using the electron test beam in a magnet field at the facility in DESY. The dE/dx resolution for ILD-TPC was estimated using this test beam data. The signal charge was measured as a function of the incident angle of the beam. In this paper, we report on the measurement of the signal charge and of the dE/dx resolution.
\end{abstract}

\begin{keyword}
Time Projection Chamber (TPC), Energy loss (dE/dx), Gas Electron Multiplier (GEM), \\Ion Back Flow (IBF), International Large Detector (ILD), International Linear Collider (ILC)
\end{keyword}

\end{frontmatter}


\section{\large Introduction}
The future ILC (International Linear Collider) consists of two linear accelerators that face each other, and which collide electrons and positrons at 250 GeV center of mass energy (The current baseline design allows for an upgrade to 1 TeV). The ILC will give precise measurement of the Higgs Boson and the top quark, and the search of new physics beyond the Standard Model.  

\par The TPC (Time Projection Chamber) will be used as a central tracker in the ILD (International Large Detector), which is one of the detector concepts for the ILC. The TPC can measure the momentum of tracks of charged particles in the magnetic field. Micro Pattern Gas Detector (MPGD) such as Gas Electron Multiplier (GEM)\cite{gem} and Micro-MEsh GAseous Structure (Micro MEGAS)\cite{megas} or the Timepix chip\cite{time} are candidate for the readout technology. Also, the TPC can reconstruct and identify particle species using energy loss (dE/dx) measured by the readout pad rows. In the reaction event of the electron-positron annihilation in the ILC experiment, it is required to identify charged particle species such as pion, kaon, electron, etc. and to reconstruct the events. For the ILD-TPC, expected dE/dx resolution is 5 \% \cite{tdr} for clear identification. 

\par
\begin{figure}[htbp]
\centering
\subfigure[Gating GEM: honeycomb structure]{
\includegraphics[width=65mm]{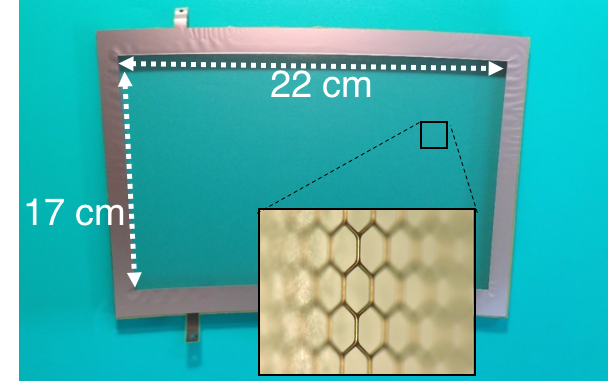}
\label{fig:gatea}}
\subfigure[Cross section]{
\includegraphics[width=55mm]{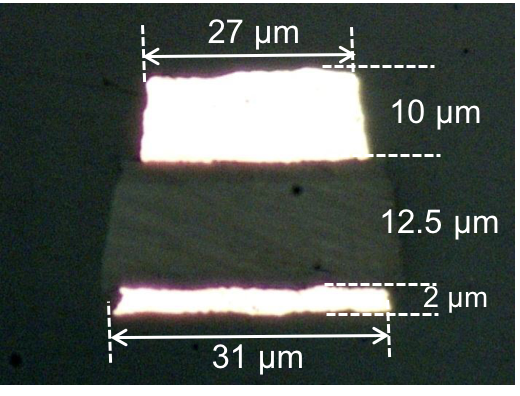}
\label{fig:gateb}}
\caption{Gating GEM: a large aperture GEM type gate device whose geometrical aperture reach 82.3 \% was manufactured cooperating with Fujikura Ltd.\cite{fuji}}
\label{fig:gating}
\end{figure}

 We performed the electron beam test of the prototype TPC with the gating device (gating GEM)\cite{gate} which will be used to suppress positive ions back flow (IBF) in a magnet field. One of ILD-TPC's problems to be solved is positive ions feedback from the gas amplification region to the drift region. IBF causes the distortion of the electric field in the drift region and the distortion of reconstructed tracks. In order to solve this problem, the gating GEM (Figure \ref{fig:gating}) is mounted on gas amplification region. At the closed state of the gating GEM it is possible to suppress IBF, at the open state the electrons from charged particle are able to pass through the gas amplification region, readout. The dE/dx resolution of our prototype TPC was measured with a potential of about 3.5V applied to the gating GEM at which the electron transmission rate was maximum at B = 0 T. Also, the dE/dx resolution was measured without the gating GEM. The dE/dx resolution of ILD-TPC (both large and small ILD models) was estimated using this beam test data.

\begin{figure}[htbp]
  \begin{center}
  \includegraphics[width=120mm]{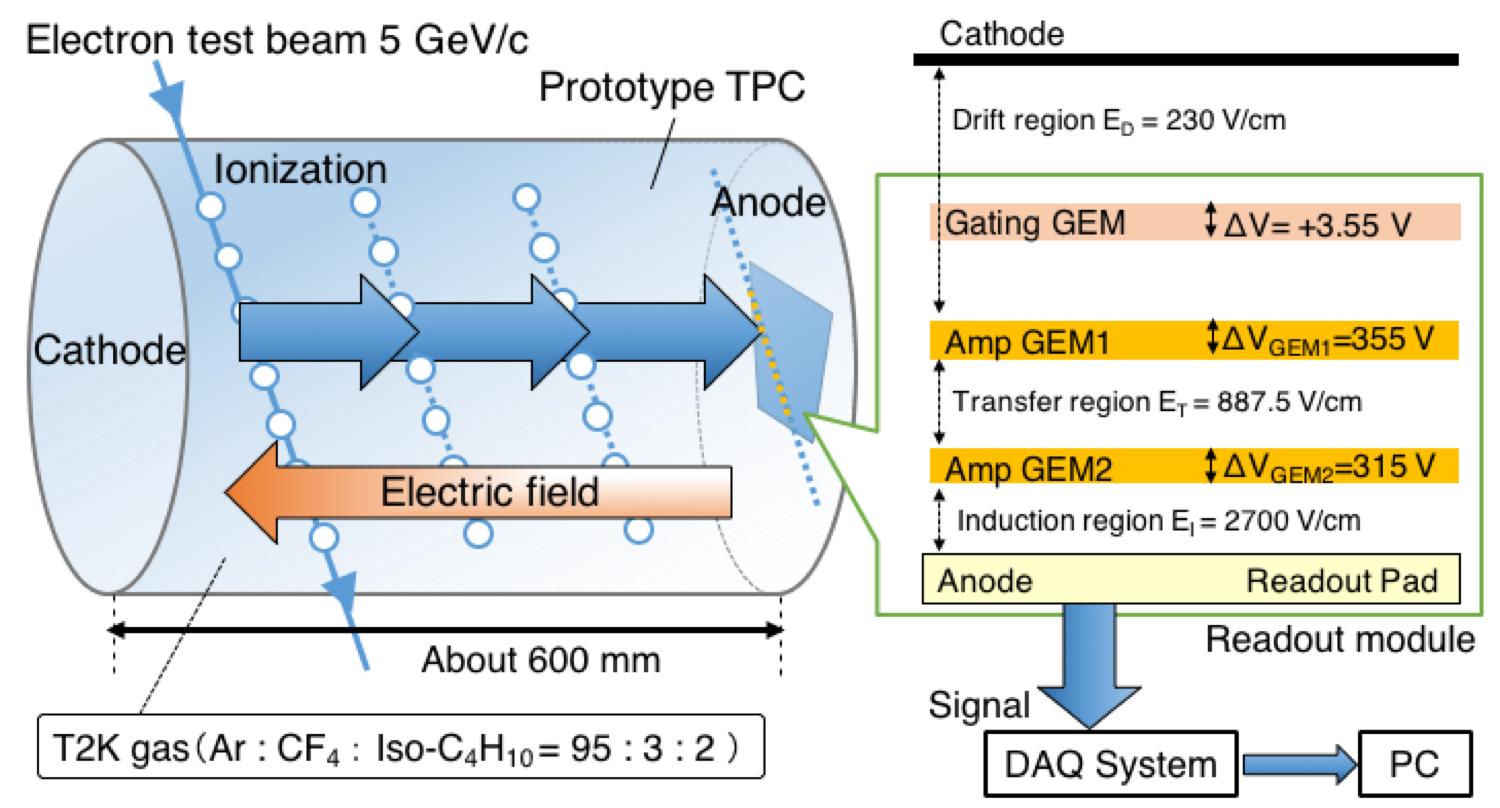} 
  \end{center}
 \caption{Experimental setup}
 \label{fig:setup}
\end{figure}

\section{\large Experimental setup}
The beam test was performed at the facility in DESY (Deutsches Elektronen-Synchrotron)\cite{desy} by the LCTPC collaboration\cite{lctpc}. Figure \ref{fig:setup} shows experimental setup. The readout module consisting of the gating GEM, amplifying GEMs and readout pads is installed into the prototype TPC. The prototype TPC is filled with T2K gas (Ar:CF${}_\text{4}$:Iso-C${}_\text{4}$H${}_\text{10}$ = 95:3:2 [\%])\cite{t2k}. The prototype TPC is installed into PCMAG (super conducting solenoid magnet) which has a magnetic field 1 T. Electric fields such as drift region, transfer region, induction region inside the prototype TPC are formed by applying a potential to the cathode and electrodes of the gating GEM and amplifying GEMs. The prototype TPC is exposed to the electron test beam (5 GeV/c) from the DESYI\hspace{-.1em}I accelerator. When the electron beam passes through the prototype TPC filled with the gas, the gas molecules are ionized. Then electrons from the ionization are generated along the tracks. The electrons drift to the anode (readout pads) due to the force by the electric field the data acquisition of their signal charge is registered by the readout electronics\cite{altro}.

\begin{figure}[h]
\centering
\subfigure[The module with gating GEM]{
\includegraphics[width=60mm]{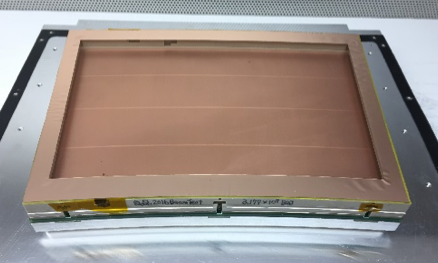}
\label{fig:wgate}}
\subfigure[The module without gating GEM]{
\includegraphics[width=60mm]{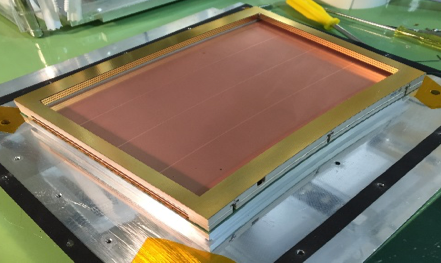}
\label{fig:field}}
\centering
\subfigure[Amplifying GEM]{
\includegraphics[width=60mm]{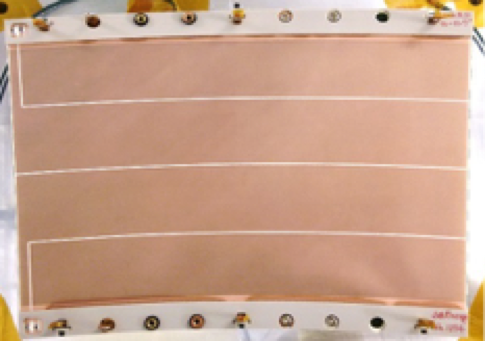}
\label{fig:gem}}
\subfigure[Readout pads]{
\includegraphics[width=60mm]{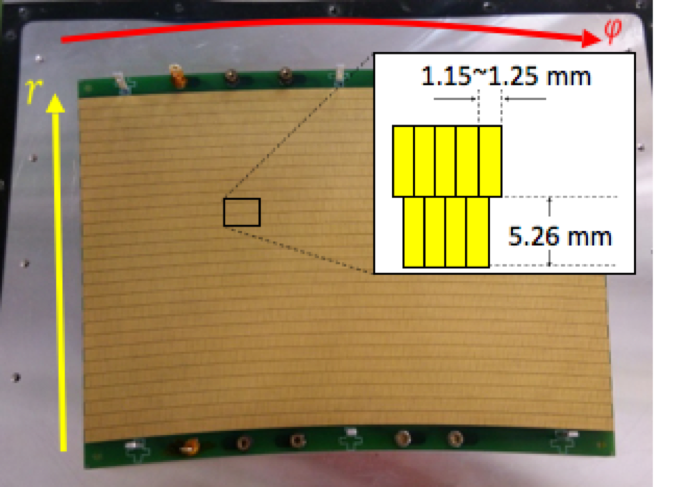}
\label{fig:pad}}
\caption{Readout module}
\label{fig:readout}
\end{figure}

\paragraph{Readout module}  The readout module consists of the gating GEM, amplifying GEMs and readout pads. We measured at the both of with or without the gating GEM (Figure \ref{fig:wgate},\ref{fig:field}). A field shaping frame was mounted on the top of the amplifying GEMs in the absence of the gating GEM (Figure \ref{fig:field}). The gating GEM is a very thin electrode foil which has insulator (polyimide) thickness of 12.5 $\mu$m and honeycomb structure (Figure \ref{fig:gating}). When two electrode sides the front and the back of the gating GEM are given a potential difference, an electric field is formed at the hexagonal holes, so it is possible to open and close the gate by changing the voltage. Two amplifying GEMs which have insulator (LCP: Liquid Crystal Polymer) thickness of 100 $\mu$m are used. The electrode of amplifying GEM is divided on one side into four parts to protect the GEM foil in case of discharge (Figure \ref{fig:gem}). Readout with 28 pad rows is used (Figure \ref{fig:pad}). 

\paragraph{Incident angle of the beam}  We measured at the beam incident angles of 0 deg and 20 deg. In the case of 20 deg, prototype TPC is turned in counterclockwise 20 deg. And we measured at drift lengths varying  from 12.5 mm to 550 mm.

\paragraph{Data analysis} At this data analysis, MarlinTPC \cite{marlin} which is the analysis software package was used. In the measurement 20,000 events data were acquired for each drift length. Tracks are reconstructed from the acquired data. Hit points on readout pad were calculated using the Center of Gravity (C.O.G). Tracks are reconstructed by Kalman Filter \cite{kalman} from hit points. In the data analysis, only tracks are selected for which the beam does not pass too close to the outer wall of the TPC.

\section{\large Measurement of dE/dx resolution}
\begin{figure}[h]
\centering
\subfigure[]{
\includegraphics[width=70mm]{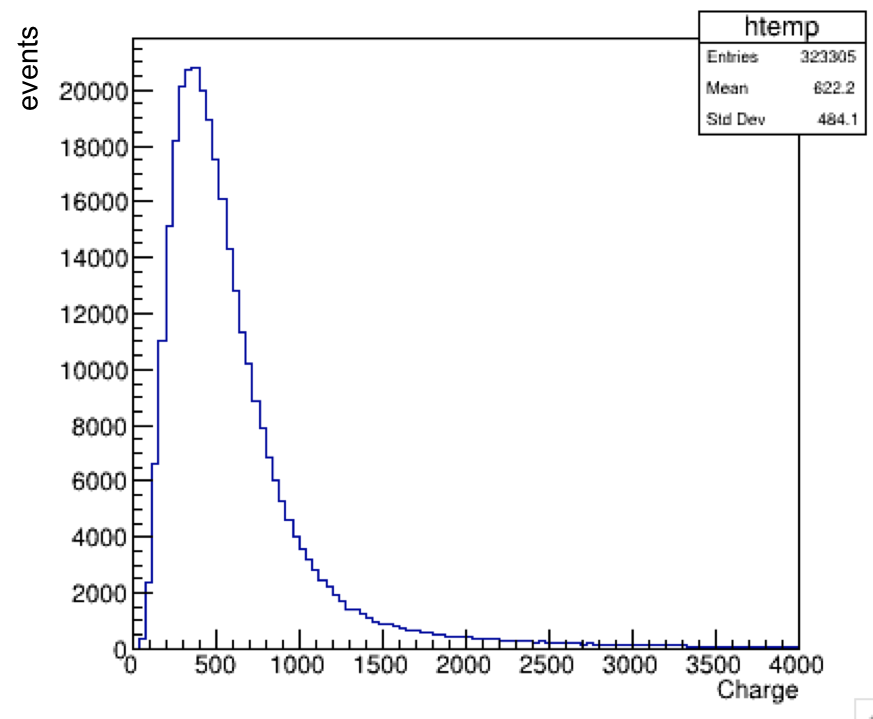}
\label{fig:landau}}
\subfigure[]{
\includegraphics[width=75mm]{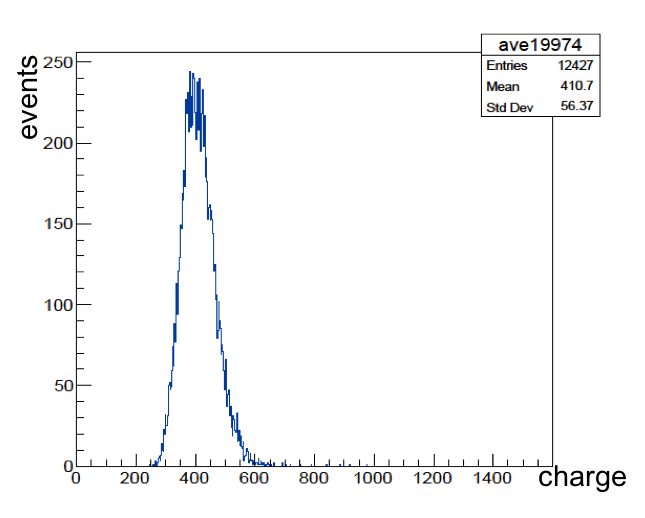}
\label{fig:truncated}}
\caption{(a)Charge distribution with gating GEM at drift length 45cm, (b)Charge distribution using truncated mean method with gating GEM at drift length 45 cm }
\label{fig:dirstiburion}
\end{figure}

The dE/dx resolution was calculated using the charges measured by 26 pad rows (out of 28), excluding the innermost and the outermost ones near the supporting frame. Figure \ref{fig:landau} shows signal charge distribution obtained at this beam test. The distribution has a typical Landau tail.  Calculating the dE/dx resolution including Landau tail part is undesirable because it causes additional fluctuation. So the dE/dx resolution is calculated using the Truncated Mean method. The pad hits were sorted by their charges and the average charge over those giving the lowest 70\% were used in order to get rid of the Landau tail. The charge of highest 30\% is truncated. Thus, the distribution such as Figure \ref{fig:truncated} is obtained. The dE/dx resolution is obtained dividing standard deviation by mean of the distribution using the Truncated Mean method.

\par
\begin{figure}[h]
\centering
\subfigure[]{
\includegraphics[width=80mm]{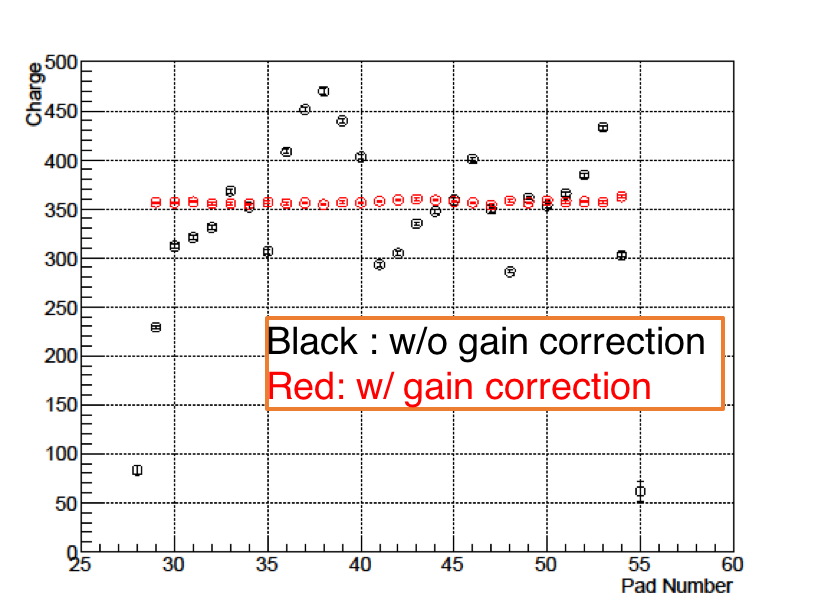}
\label{fig:gainv}}
\subfigure[]{
\includegraphics[width=75mm]{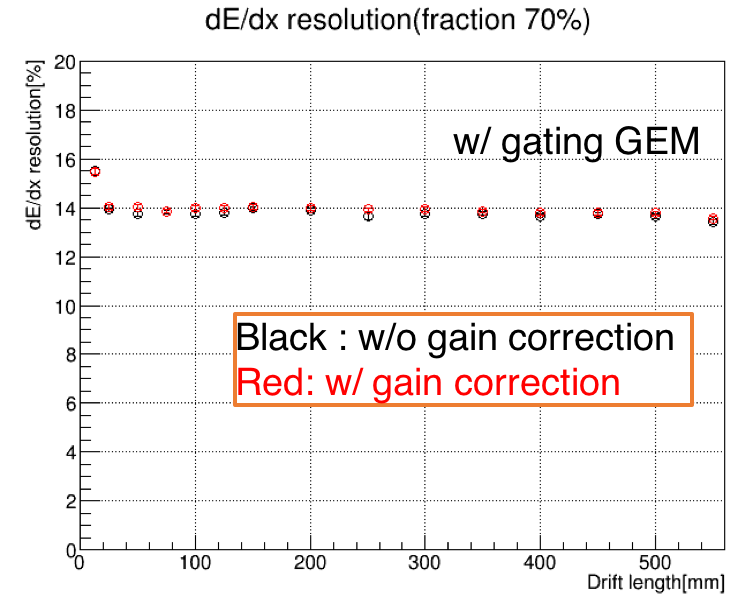}
\label{fig:gainc}}
\caption{(a)Pad row dependence of charge with gating GEM at drift length 45cm, (b)dE/dx resolution with or without gain correction }
\label{fig:gaincorr}
\end{figure}

 The signal charge has been corrected for the pad-row to pad-row gain variation before sorting for getting the truncated mean. Each charge is corrected by dividing by a factor which is the average of the charge of all pads of a given pad row. Figure \ref{fig:gainv} shows the pad row dependence of the charge. Black points in figure \ref{fig:gainv} are the gain fluctuation due to non-uniformity of the amplifying GEM's thickness and the structure of the amplifying GEM's divided electrode. Red points in figure \ref{fig:gainv} shows the variation after applying the correction factor. Figure \ref{fig:gainc} is the result of the dE/dx resolution with or without gain correction. The dE/dx resolution averaged  over drift length without gain correction is 13.96 $\pm$ 0.02 \%. The dE/dx resolution averaged over drift length with gain correction is 13.87 $\pm$ 0.02 \%. The dE/dx resolution is rather insensitive to the pad-row to pad-row gain variation as expected.

\begin{figure}[h]
\centering
\subfigure[]{
\includegraphics[width=85mm]{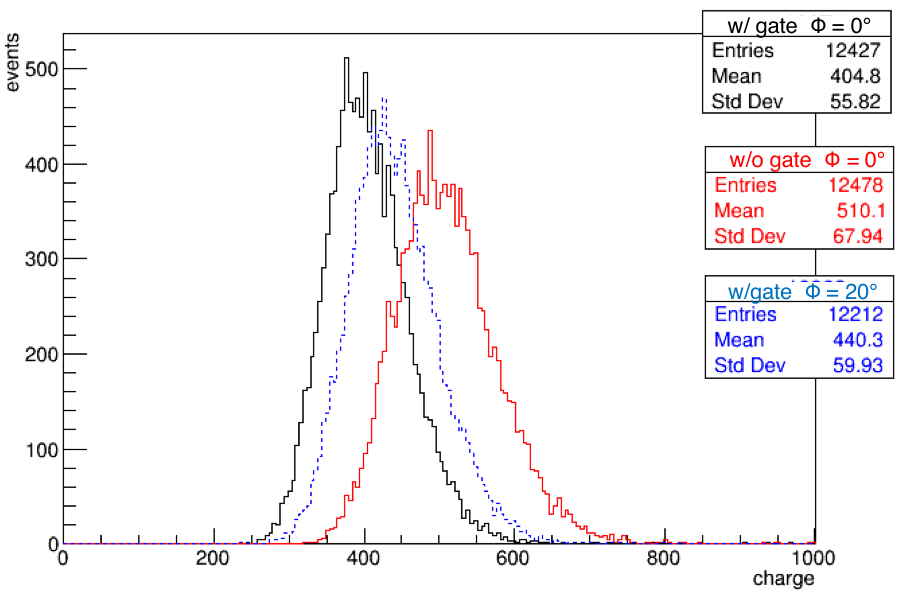}
\label{fig:tru_dist}}
\subfigure[]{
\includegraphics[width=75mm]{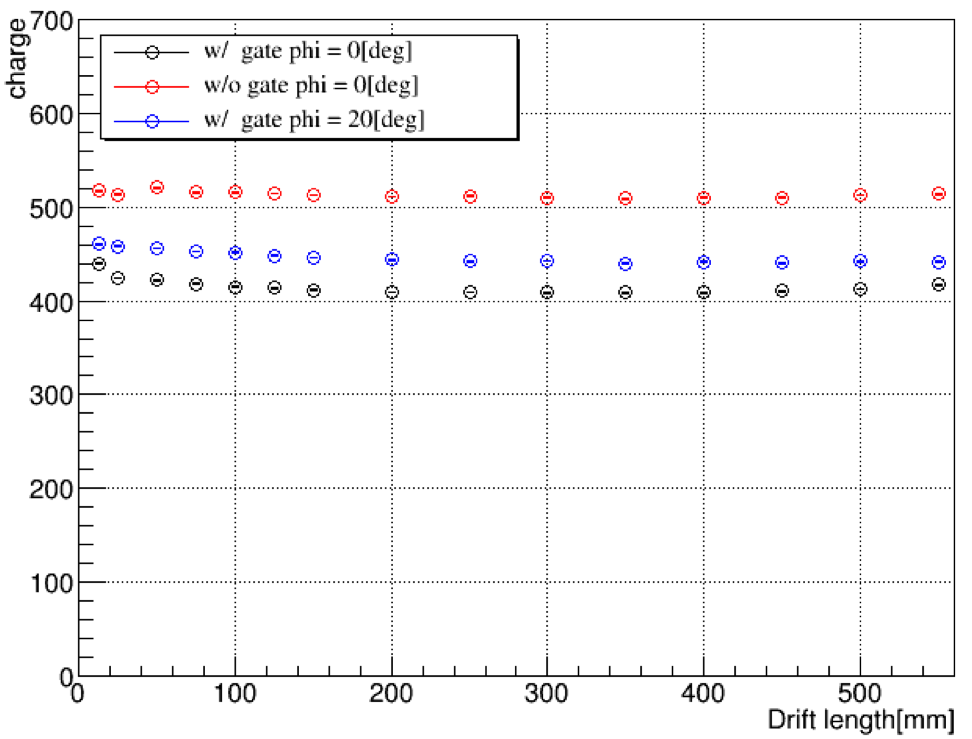}
\label{fig:tru_drift}}
\caption{(a)Charge distribution using Truncated Mean method at fraction of 70 \%, (b)Signal charge (Truncated Mean) v.s. drift length  }
\label{fig:charge}
\end{figure}

\paragraph{Signal charge (Truncated Mean)} Figure \ref{fig:tru_dist}shows an example of the charge distribution using the Truncated Mean method. Figure \ref{fig:tru_drift} is a graph in which the vertical axis is mean of the charge distribution using the Truncated Mean method and the horizontal axis is the drift length. Black line in figure \ref{fig:charge} is with gating GEM, red line in figure \ref{fig:charge} is without gating GEM. In the case of with the gating GEM, the signal charge is less than without the gating GEM because there are some electrons adsorbed to the electrode of the gating GEM. The blue dash line in figure \ref{fig:charge} shows the data with gating GEM at beam incident angle is 20 deg. In this case, the signal charge increases because the track becomes longer. 

\begin{figure}[htbp]
  \begin{center}
  \includegraphics[width=90mm]{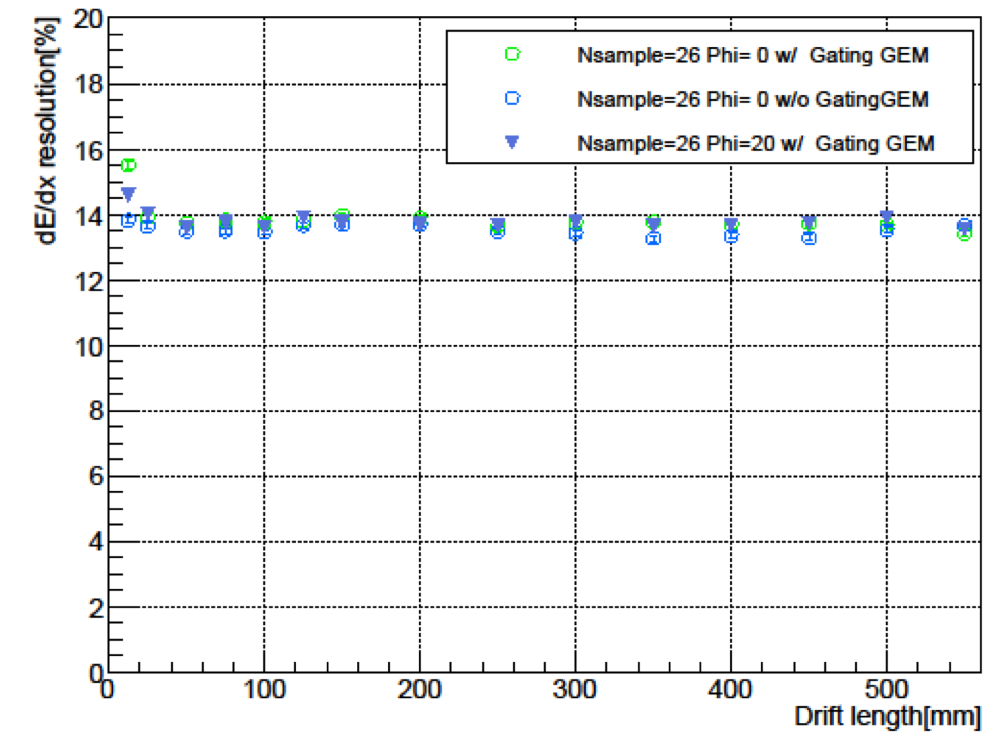} 
  \end{center}
 \caption{The dE/dx resolution of prototype TPC (1 readout module: 26 pad rows)}
 \label{fig:proto}
\end{figure}

\paragraph{dE/dx resolution of prototype TPC (1 readout module: 26 layers)} The dE/dx resolution is calculated from mean and standard deviation of the charge distribution using the Truncated Mean method. Figure \ref{fig:proto} shows result of the dE/dx resolution of prototype TPC (1 readout module). The dE/dx resolution at shortest drift length is worse because some track length may be lost for pieces of track leaving the drift volume. The average of the dE/dx resolution of prototype TPC is calculated for drift lengths of greater than 12.5mm (the same cut is used below). The dE/dx resolution averaged over drift length of prototype TPC with the gating GEM is 13.80 $\pm$ 0.02 \%. In the prototype TPC without the gating GEM, the dE/dx resolution averaged over drift length is 13.52 $\pm$ 0.02 \%. In the prototype TPC with the gating GEM at beam incident angle is 20deg, the dE/dx resolution averaged over drift length is 13.70 $\pm$ 0.02 \%. The dE/dx resolution is rather insensitive to the presence or absence of the gating GEM.
 
\section{\large Extrapolation of dE/dx resolution of ILD-TPC}
\begin{figure}[htbp]
  \begin{center}
  \includegraphics[width=120mm]{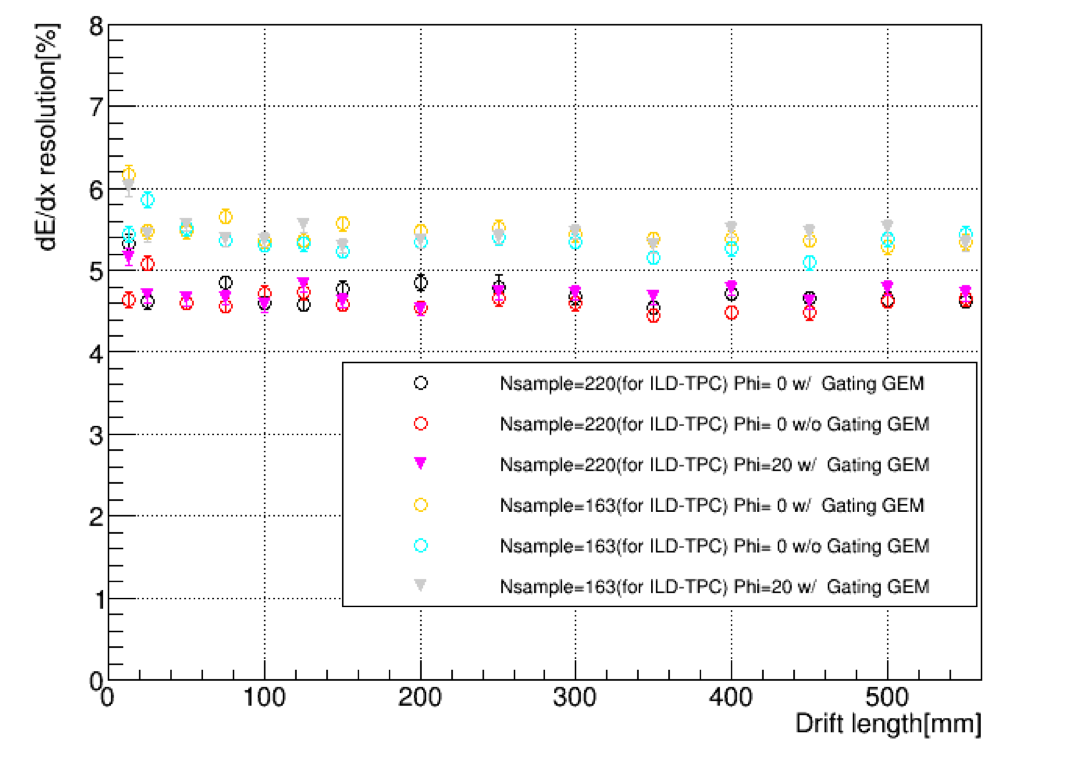} 
  \end{center}
 \caption{Extrapolation of the dE/dx resolution of ILD-TPC (both models: large model-TPC has 220 pad rows, small model-TPC has 163 pad rows)}
 \label{fig:ildtpc}
\end{figure}

Although the dE/dx resolution of prototype TPC (1 module) is obtained using the charges measured by 26 pad rows, large model of ILD-TPC has 220 pad rows\cite{tdr}, and small model of ILD-TPC has 163 pad rows. For the estimation of the dE/dx resolution of the large model of ILD-TPC, tracks with 220 hit pad rows were simulated by combining the charge measurements in 9 events (tracks). Figure \ref{fig:ildtpc} shows the result of the dE/dx resolution of the large model of ILD-TPC. The dE/dx resolution averaged over drift length of large model of ILD-TPC with the gating GEM is 4.66 $\pm$ 0.02 \%. In the large model-TPC without the gating GEM, the dE/dx resolution averaged over drift length is 4.61 $\pm$ 0.02 \%. In large model-TPC with the gating GEM at beam incident angle is 20deg, the dE/dx resolution averaged over drift length is 4.68 $\pm$ 0.02 \%. The dE/dx resolution is rather insensitive to the presence or absence of the gating GEM. Also, the dE/dx resolution of small model of ILD-TPC was estimated in similar way using 7 events (tracks) and is seen in Figure \ref{fig:ildtpc}. The dE/dx resolution averaged over drift length of small model of ILD-TPC with the gating GEM is 5.46 $\pm$ 0.02 \%. In the small model-TPC without the gating GEM, the dE/dx resolution averaged over drift length is 5.35 $\pm$ 0.02 \%. In the small model-TPC with the gating GEM at beam incident angle is 20deg, the dE/dx resolution averaged over drift length is 5.42 $\pm$ 0.02 \%. The dE/dx resolution of large model-TPC is better than the small model-TPC. 

\section{\large Summary}
We performed the electron beam test of the prototype TPC with the gating device (gating GEM) which is used to suppress positive ions back flow (IBF). The dE/dx resolution was measured in this beam test. The dE/dx resolution of ILD-TPC (both large and small ILD models) was estimated. The dE/dx resolution is rather insensitive to the pad-row to pad-row gain variation, as well as to the number of primary electrons, i.e. to the presence or absence of the gating GEM. The dE/dx resolution of the ILD-TPC (large-model) with a gating GEM was estimated to be about 4.7 \%  for 5 GeV/c electrons on the Fermi plateau. In the small model-TPC with the gating GEM, the dE/dx resolution was estimated to be about 5.5 \%.

\section*{\large Acknowledgements}
This study was supported by the LCTPC collaboration. The production of the gating GEM is cooperative effort by Fujikura Ltd. The author would like to thank the members of the LCTPC group and Fujikura Ltd. The measurements leading to these results have been performed at the Test Beam Facility at DESY Hamburg (Germany), a member of the Helmholtz Association.

\bibliographystyle{unsrt}

\end{document}